\documentclass[prd,aps,floats,twocolumn]{revtex4}

\usepackage[dvips]{graphicx}
\usepackage{amssymb}
\usepackage{amsmath}

\usepackage{color}

\begin{document}

\title{Binary black hole merger: symmetry and the spin expansion}

\author{Latham Boyle, Michael Kesden and Samaya Nissanke}

\affiliation{Canadian Institute for Theoretical Astrophysics
  (CITA), University of Toronto, 60 St.~George Street, Toronto,
  Ontario, M5S 3H8, Canada}

\date{September 2007}
                            
\begin{abstract}
  We regard binary black hole (BBH) merger as a map from a simple
  initial state (two Kerr black holes, with dimensionless spins ${\bf
    a}$ and ${\bf b}$) to a simple final state (a Kerr black hole with
  mass $m$, dimensionless spin ${\bf s}$, and kick velocity ${\bf
    k}$).  By expanding this map around ${\bf a} = {\bf b} = 0$ and
  applying symmetry constraints, we obtain a simple formalism that is
  remarkably successful at explaining existing BBH simulations.  It
  also makes detailed predictions and suggests a more efficient way of
  mapping the parameter space of binary black hole merger.  Since we
  rely on symmetry rather than dynamics, our expansion complements
  previous analytical techniques.
\end{abstract}
\maketitle 

In binary black hole (BBH) merger, two black holes $A$ and $B$ (with
masses $M_{a}$, $M_{b}$, and spins ${\bf a}$, ${\bf b}$) inspiral due
to the emission of gravitational radiation and eventually merge to
form a final black hole with mass $m$, spin vector ${\bf s}$, and
recoil (or ``kick'') velocity ${\bf k}$.  How do the final quantities
$\{m,{\bf k},{\bf s}\}$ depend on the initial quantities
$\{M_{a},M_{b},{\bf a},{\bf b}\}$?  This is a classic problem in
general relativity (GR), with important implications for astrophysics,
cosmology and gravitational wave (GW) detection.  For example, when
two galaxies merge, their central supermassive black holes also merge.
The final quantities $\{m,{\bf k},{\bf s}\}$ from these supermassive
BBH mergers are linked (see \cite{Pretorius:2007nq} and references
therein) to a variety of astrophysical observables including: (i) the
quasar luminosity function; (ii) the location of quasars relative to
their host galaxies; (iii) the orientation and shape of jets in active
galactic nuclei; (iv) the correlation between black hole mass and
velocity dispersion in the surrounding stellar bulge; (v) the density
profile in galactic centers.  The quantities $\{m,{\bf k},{\bf s}\}$
are also intimately related to the spectrum of quasi-normal ringdown
modes after BBH merger --- a key observable for probing black holes
and strong-field GR with GW detectors.

Following recent numerical breakthroughs \cite{Pretorius:2005gq,
  Campanelli:2005dd, Baker:2005vv}, a number of groups can now
simulate entire BBH mergers.  In particular, they can choose a set of
initial quantities $\{M_{a},M_{b},{\bf a},{\bf b}\}$ and compute the
corresponding final quantities $\{m,{\bf k},{\bf s}\}$.  As more of
these extremely time-intensive simulations have gradually accumulated,
certains patterns and trends have emerged.  In previous work, some of
these patterns have been described by empirical fitting functions
\cite{Campanelli:2007ew, Baker:2007gi} which are loosely inspired by
(but not derived from) post-Newtonian formulae \cite{Kidder:1995zr}.
In this paper, we show how these same patterns (and others) may in
fact be derived from elementary symmetry arguments.  This perspective
has several advantages which make it complementary to post-Newtonian
and numerical techniques.  As we shall explain, the resulting
formalism: (i) provides a simple conceptual understanding of BBH
merger, accessible to non-experts; (ii) makes a host of new and
derived predictions which go beyond the fitting formulae
\cite{Campanelli:2007ew, Baker:2007gi}; (iii) suggests an efficient
way to map out the parameter space of BBH mergers with simulations;
and (iv) provides a map $\{M_{a},M_{b},{\bf a},{\bf
  b}\}\!\to\!\{m,{\bf k},{\bf s}\}$ which is useful for astrophysical
applications (including semi-analytic models or N-body simulations of
black hole growth in galaxy mergers and dense stellar clusters) which
wish to include BBH mergers, but cannot hope to follow the detailed
merger dynamics.

Although the merger process involves complicated non-linear dynamics,
the initial and final states of the system are rather simple and
symmetric.  The initial state consists of two widely separated Kerr
black holes, and the final state is a single Kerr black hole.  The
idea of this {\it Letter} is to see how much we can learn by
considering the merger process as a map between these simple initial
and final states, ignoring as much as possible the detailed dynamics
in between.  Under three operations (rotation ``$R$,'' parity ``$P$,''
and exchange ``$X$''), the initial and final states transform
according to simple rules.  The dependence of the final state on the
initial state is constrained to be consistent with these rules.  We
present a naive formalism based on systematically applying these
symmetry considerations to a well chosen Taylor expansion.
\begin{figure}
  \begin{center}
    \includegraphics[width=2.9in]{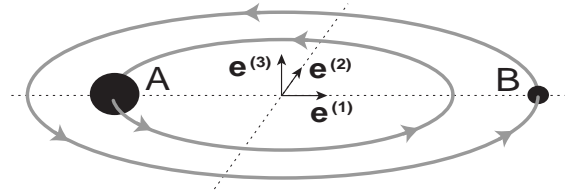}
  \end{center}
  \caption{The orthonormal triad presented in the text.}
  \label{triadfig}
\end{figure}

Imagine two black holes, $A$ and $B$, in a circular orbit
\footnote{For most astrophysically relevant systems, elliptical BBH
  orbits are expected to circularize long before merger
  \cite{Peters:1963ux}.  The elliptical case is treated in
  \cite{Boyle:2007ru}.}.  The orbit gradually shrinks due to
gravitational-wave emission, until $A$ and $B$ eventually merge.
Consider an initial instant when the holes are far apart --- far
enough that they may be approximated as two Kerr black holes with well
defined masses ($M_{a}$ and $M_{b}$) and dimensionless spins (${\bf
  a}\equiv{\bf S}_{a}/M_{a}^{2}$ and ${\bf b}\equiv{\bf S}_{b}
/M_{b}^{2}$), orbiting in a well defined plane that is perpendicular
to the initial (dimensionless) orbital angular momentum ${\bf
  L}_{0}/M^{2}$ (where $M=M_{a}+M_{b}$, and $G_{N}^{}=c=1$).  Long
after the merger is complete, the gravitational radiation has total
energy $E_{{\rm rad}}$ and total angular momentum ${\bf J}_{{\rm
    rad}}$; and the final Kerr black hole has dimensionless mass
$m=M_{f}/M$, dimensionless kick velocity ${\bf k}$ (relative to the
center of mass), and dimensionless spin ${\bf s}={\bf S}_{f}
/M_{f}^{2}$.

At the initial instant mentioned above, we define an orthonormal triad
$\{{\bf e}_{}^{(1)}, {\bf e}_{}^{(2)},{\bf e}_{}^{(3)}\}$ as shown in
Fig.~\ref{triadfig}: ${\bf e}_{}^{(3)}$ is the direction of the
orbital angular momentum, ${\bf e}_{}^{(1)}$ is the direction from $A$
to $B$, and ${\bf e}_{}^{(2)}={\bf e}_{}^{(3)}\times{\bf e}_{}^{(1)}$.
Then the circular BBH's initial state is specified by 7 numbers: the
mass ratio $q$ and the spin components
\begin{equation}
  \label{ab_components}
  a_{i}^{}={\bf a}\cdot{\bf e}_{}^{(i)}\qquad\qquad
  b_{i}^{}={\bf b}\cdot{\bf e}_{}^{(i)}
\end{equation}
relative to the orthonormal triad.  Similarly, let us convert the
final vectors into their separate triad components
\begin{equation}
  \label{ks_components}
  k_{i}^{}={\bf k}\cdot{\bf e}_{}^{(i)}\qquad\qquad
  s_{i}^{}={\bf s}\cdot{\bf e}_{}^{(i)}.
\end{equation}
If we apply a global 3-dimensional rotation $R$ to the entire binary
system (as if it were a single rigid body), initial and final
quantities like ${\bf a}$, ${\bf b}$, ${\bf k}$, and ${\bf s}$ rotate
as vectors --- as do the triad elements ${\bf e}_{}^{(i)}$.
Therefore, the corresponding triad components ($a_{i}$, $b_{i}$,
$k_{i}$, and $s_{i}$) transform as {\it scalars} under $R$.  By
working with triad components, all of our subsequent formulae are
manifestly consistent with rigid 3-dimensional rotations $R$.

We can view any final quantity $f$ (such as $m$, $s_{i}$, or $k_{i}$)
as a function of the initial quantities:
\begin{equation}
  \label{f_func}
  f=f(q,a_{1},a_{2},a_{3},b_{1},b_{2},b_{3}).
\end{equation}
Let us Taylor expand this function around ${\bf a}={\bf b}=0$:
\begin{equation}
  \label{f_series}
  f=f_{}^{m_{1}m_{2}m_{3}|n_{1}n_{2}n_{3}}(q)a_{1}^{m_{1}}
  a_{2}^{m_{2}}a_{3}^{m_{3}}b_{1}^{n_{1}}b_{2}^{n_{2}}b_{3}^{n_{3}}.
\end{equation}
Since a Kerr black hole has maximum spin ${\bf S}_{i}\leq M_{i}^{2}$,
$|{\bf a}|$ and $|{\bf b}|$ are both $\leq 1$ and it is not
unreasonable to hope that the Taylor series might be convergent over
most or even all of this range of initial spins.

We will now use additional symmetries to restrict the coefficients in
this expansion.  First consider a parity transformation $P$ that
reflects every point of the binary system through the origin (the
center of mass).  Under $P$, the mass ratio $q$ is unchanged, while
the triad components transform as
\begin{equation}
  \label{Ptrans}
  \begin{array}{lcl}
    \{a_{1},a_{2},a_{3}\}&\to&\{-a_{1},-a_{2},a_{3}\} \\
    \{b_{\;\!1},b_{\;\!2},b_{\;\!3}\}&\to&
    \{-b_{\;\!1},-b_{\;\!2},b_{\;\!3}\} 
  \end{array}.
\end{equation}
Thus, each term on the right-hand-side of Eq.~(\ref{f_series}) picks
up a factor of $(-1)^{\gamma}$, where
\begin{equation}
  \gamma=m_{1}^{}+m_{2}^{}+n_{1}^{}+n_{2}^{}.
\end{equation}
If $f$ transforms under $P$ as $f\to(\pm)_{P}^{}\,f$, the coefficients
in Eq.~(\ref{f_series}) must satisfy the constraint
\begin{equation}
  \label{P_constraint}
  f_{}^{m_{1}m_{2}m_{3}|n_{1}n_{2}n_{3}}(q)=(\pm)_{P}^{}(-1)^{\gamma}
  f_{}^{m_{1}m_{2}m_{3}|n_{1}n_{2}n_{3}}(q).
\end{equation}
In other words, if $f$ is even (odd) under $P$, then
$f_{}^{m_{1}m_{2}m_{3} |n_{1}n_{2}n_{3}}(q)$ must vanish when $\gamma$
is odd (even).

Finally, we apply an ``exchange transformation'' $X$.  This leaves the
physical system absolutely unchanged, and simply swaps the labels of
the two black holes, $A\leftrightarrow B$.  Under $X$, the mass ratio
transforms as $q\to1/q$, while the triad components transform as
\begin{equation}
  \label{Xtrans}
  \begin{array}{lcl}
    \{a_{1},a_{2},a_{3}\}&\to&\{-b_{\;\!1},-b_{\;\!2},b_{\;\!3}\} \\
    \{b_{\;\!1},b_{\;\!2},b_{\;\!3}\}&\to&\{-a_{1},-a_{2},a_{3}\}
  \end{array}.
\end{equation}
If $f$ transforms under $X$ as $f\to(\pm)_{X}^{}\,f$, the
coefficients in Eq.~(\ref{f_series}) must satisfy the constraint
\begin{equation}
  \label{X_constraint}
  f_{}^{m_{1}m_{2}m_{3}|n_{1}n_{2}n_{3}}(q)=(\pm)_{X}^{}
  (-1)_{}^{\gamma}f_{}^{n_{1}n_{2}n_{3}|m_{1}m_{2}m_{3}}
  (1/q).
\end{equation}
Equivalently, and more conveniently, if $f$ transforms under $PX$ ($P$
followed by $X$, or vice versa) as $f\to(\pm)_{PX}^{}\,f$, the
coefficients in Eq.~(\ref{f_series}) must satisfy the constraint
\begin{equation}
  \label{PX_constraint}
  f_{}^{m_{1}m_{2}m_{3}|n_{1}n_{2}n_{3}}(q)=(\pm)_{PX}^{}
  f_{}^{n_{1}n_{2}n_{3}|m_{1}m_{2}m_{3}}(1/q).
\end{equation}
The transformation laws under $P$, $X$, and $PX$, for various final
quantities $f$, are summarized in Table~\ref{transformation_table}.
\begin{table}
  \begin{center}
    \begin{tabular}{|l|c|c|c|c|}
      \hline
      $f$ & $m$, $E_{}^{{\rm rad}}$, $s_{3}$, $J_{3}^{{\rm rad}}$ & 
      $s_{1}^{}$, $s_{2}^{}$, $J_{1}^{{\rm rad}}$,
      $J_{2}^{{\rm rad}}$ &
      $k_{1}^{}$, $k_{2}^{}$ & $k_{3}^{}$ \\
      \hline
      $(\pm)_{P}^{}$ & $+$ & $-$ & $+$ & $-$ \\
      \hline
      $(\pm)_{X}^{}$ & $+$ & $-$ & $-$ & $+$ \\
      \hline
      $(\pm)_{PX}^{}$ & $+$ & $+$ & $-$ & $-$ \\
      \hline
    \end{tabular}
  \end{center}
  \caption{Transformation under $P$, $X$, and $PX$,
    for various final quantities $f$.}
  \label{transformation_table}
\end{table}

Eqs.~(\ref{f_series}, \ref{P_constraint}, \ref{PX_constraint}) imply a
number of results that are {\it exact} ({\it i.e.}\ valid to all
order in the Taylor expansion).  For example, if the initial spin
configuration has ${\bf a}\propto {\bf b}\propto{\bf e}_{}^{(3)}$,
then $\gamma=0$ and parity requires ${\bf s}\propto{\bf e}_{3}$ and
$k_{3}=0$.  In the equal-mass case, $q=1=1/q$ and
Eq.~(\ref{PX_constraint}) demands that all final quantities odd under
$PX$ (like $k_{i}$) vanish for ${\bf a}={\bf b}$.  Again in the
equal-mass case, if the initial spin configuration satisfies
$(a_{1},a_{2},a_{3})= (-b_{1},-b_{2},b_{3})$, then Eq.~(\ref{Xtrans})
becomes the identity mapping and all quantities odd under $X$ (like
$s_{1},s_{2},k_{1},k_{2}$) must vanish.

Although these exact results are interesting, the real power of our
formalism lies in the many {\it approximate} predictions that it
makes.  To illustrate this, consider the quantity $k_{1}$.  From
Table~\ref{transformation_table}, $k_{1}$ has $(\pm)_{P}^{}=+1$ and
$(\pm)_{PX}^{}=-1$.  Start from the general Taylor expansion:
\begin{equation}
  \label{k1_1st_order}
  k_{1}^{}=k_{1}^{m_{1}m_{2}m_{3}|n_{1}n_{2}n_{3}}a_{1}^{m_{1}}
  a_{2}^{m_{2}}a_{3}^{m_{3}}b_{1}^{n_{1}}b_{2}^{n_{2}}b_{3}^{n_{3}}.
\end{equation}
The zeroth-order term $k_{1}^{000|000}$ vanishes by
Eq.~(\ref{PX_constraint}).  At first order, the Taylor expansion has
six terms (one for each of the spin components $\{a_{1},a_{2},
a_{3},b_{1},b_{2},b_{3}\}$).  But constraints (\ref{P_constraint}) and
(\ref{PX_constraint}) cut this down to a single term:
\begin{equation}
  k_{1}^{}=k_{1}^{001|000}(a_{3}-b_{3}).
\end{equation}
Thus, at leading order, we predict $k_{1}^{}\propto
(a_{3}^{}\!-\!b_{3}^{})$, a trend seen in simulations.  But we also
predict that this leading-order behavior should be corrected by
quadratic terms of a specific form.  Although the naive Taylor
expansion contains $21$ new terms at 2nd order in spin,
Eqs.~(\ref{P_constraint}) and (\ref{PX_constraint}) reduce this to
only 5 new terms, yielding the general 2nd-order formula
\begin{eqnarray}
  \label{k1_2nd_order}
  k_{1}^{}\!&\!=\!&\!k_{1}^{001|000}(a_{3}\!-\!b_{3})\!+\!
  k_{1}^{002|000}(a_{3}^{2}\!-\!b_{3}^{2})\nonumber \\
  \!&\!+\!&\!k_{1}^{200|000}(a_{1}^{2}\!-\!b_{1}^{2})
  \!+\!k_{1}^{110|000}(a_{1}a_{2}\!-\!b_{1}b_{2})\nonumber\\ 
  \!&\!+\!&\!k_{1}^{020|000}(a_{2}^{2}\!-\!b_{2}^{2})
  \!+\!k_{1}^{100|010}(a_{1}^{}b_{2}^{}\!-\!b_{1}^{}a_{2}^{}).\quad
\end{eqnarray}  
This expansion could be continued to 3rd order and beyond.  At each
order, we obtain more terms --- although far fewer than a naive Taylor
expansion would suggest.
\begin{figure}
  \begin{center}
    \includegraphics[width=3.1in]{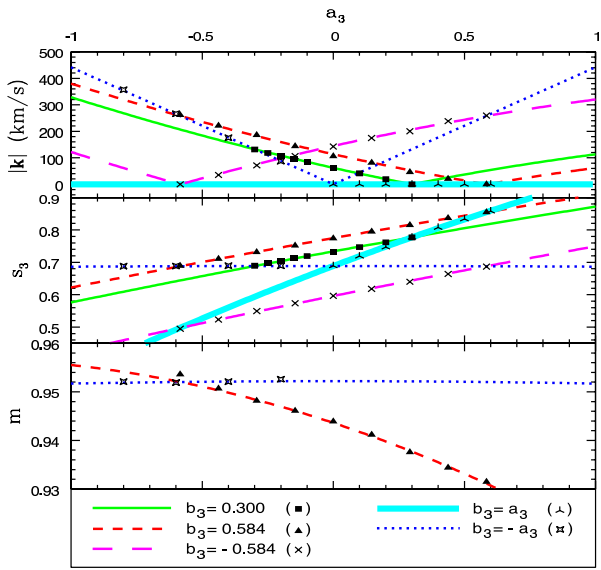}
  \end{center}
  \caption{Equal-mass BBHs with ${\bf a}\propto{\bf b}\propto
    {\bf e}_{}^{(3)}$.  Final kick $|{\bf k}|$, spin $s_{3}$, and mass
    $m$ are plotted versus $a_{3}$.  The five curves in the top two
    panels correspond to the five simulation sequences in
    \cite{Rezzolla:2007xa}; within each sequence $b_{3}$ takes on the
    values shown in the bottom panel as $a_{3}$ is varied.  For
    presentation purposes we have exchanged $a_{3}$ and $b_{3}$ for
    the case $b_{3} = -0.584$.  Cross and triangle data points for the
    mass $m$ are taken from \cite{Herrmann:2007ac} and
    \cite{Pollney:2007ss}, respectively.}
  \label{Case1fig}
\end{figure}

Our symmetry arguments are unable to determine the numerical values of
the 6 coefficients ($k_{1}^{001|000}$, etc.) in
Eq.~(\ref{k1_2nd_order}).  These coefficients must be calibrated by 6
equal-mass simulations with independent initial spin configurations.
If each simulation measures a reliable value for $k_{1}$, we can
simply {\it solve} for these coefficients.  Each of these simulations
should be performed at a fixed value of some inspiral parameter (such
as the orbital separation $r_{0}$ or angular momentum $L_{0}$) which
monotonically varies as the orbit shrinks.  Having determined the
coefficients, one can {\it predict} (with 2nd-order accuracy) the
value of $k_{1}$ resulting from {\it any} arbitrary configuration of
the initial spins at the {\it same} fixed value of the inspiral
parameter.  Post-Newtonian methods \cite{Kidder:1995zr} can then
relate coefficients determined at other fixed inspiral parameter
values \cite{Boyle:2007ru}.

Similar arguments to those given in the $k_{1}^{}$ example apply to
any final quantity $f$, including the final mass $m$, and the
magnitude and components of ${\bf k}$ and ${\bf s}$.

The explanatory power of this ``spin expansion'' formalism is
illustrated by a few simple examples.  (For more details, we refer the
reader to a subsequent paper \cite{Boyle:2007ru}, in which we discuss
the formalism's new predictions in more depth, and test them in detail
against currently available simulations.)  We begin with the case of
equal-mass BBHs with spins aligned or anti-aligned with ${\bf L}_{0}$
({\it i.e.}\ ${\bf a}\propto{\bf b}\propto{\bf e}_{}^{(3)}$).
Expanding the final kick $k_{i}$ to 2nd-order in initial spins yields
3 terms for $|{\bf k}|^{2}$, and 4 terms each for the final spin
$s_{3}$, and mass $m$ \footnote{A few days before our {\it Letter}
  appeared, \cite{Rezzolla:2007xa} gave a related expansion for this
  special case ${\bf a}\propto{\bf b}\propto{\bf e}_{}^{(3)}$.}.
Fig.~\ref{Case1fig} shows our best fits for this configuration.  As
seen in the figure, the data are well-described by the linear terms in
the spin expansion, and the fits also show evidence for small second
order corrections of the predicted form.

\begin{figure}
  \begin{center}
    \includegraphics[width=3.1in]{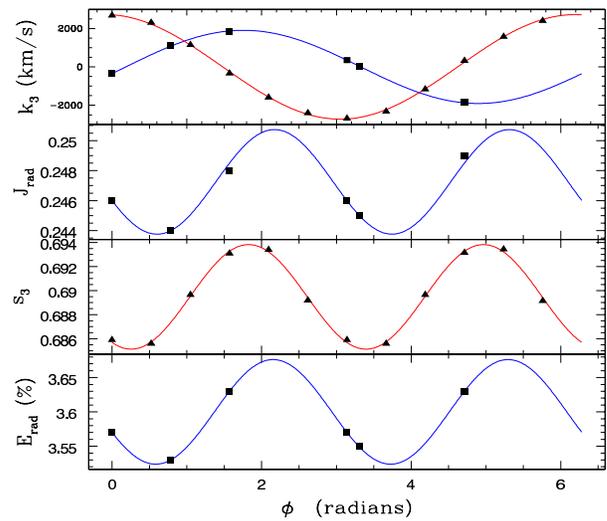}
  \end{center}
  \caption{Equal-mass BBHs with equal and opposite spins lying in the
    orbital plane.  Kick velocity $k_3$, radiated angular momentum
    $J_{\rm rad}$, final spin $s_3$, and percentage of radiated energy
    $E_{\rm rad}$ are plotted against the angle $\phi$ between the
    initial spin ${\bf a}$ and ${\bf e}^{(1)}$.  The blue curves show
    fits to the square data points taken from \cite{Campanelli:2007cg},
    while the red curves show fits to the triangle data points of
    \cite{Bruegmann:2007zj}.}
  \label{Case2fig}
\end{figure}

Fig.~\ref{Case2fig} shows our fits for the ``superkick'' configuration
(equal-mass BBHs with equal and opposite spins and
$a_{i}\!=\!-b_{i}\!=\!a(\cos\phi,\sin\phi,0).$ The leading-order terms
in the spin expansion explain the previously noticed
\cite{Campanelli:2007cg,Bruegmann:2007zj} behavior $k_{3}=A a
\cos(\phi-\phi_1)$ where $A$ and $\phi_{1}$ are constants (top panel
in Fig.~\ref{Case2fig}).  Keeping terms to 2nd-order, the spin
expansion also correctly predicts that the final quantities
$f=\{J_{{\rm rad}}, s_{3}, E_{{\rm rad}}\}$ all behave as $f=B+C
a^{2}\cos(2\phi-\phi_2)$, where $B$, $C$, and $\phi_2$ are constants
(bottom 3 panels in Fig.~\ref{Case2fig}).  This $\cos(2\phi-\phi_2)$
behavior highlights the power of the spin expansion to go beyond
previous linear post-Newtonian fitting formulae to uncover and {\it
  explain} new and essentially non-linear behavior in the simulations.

\begin{figure}
  \begin{center}
    \includegraphics[width=3.1in]{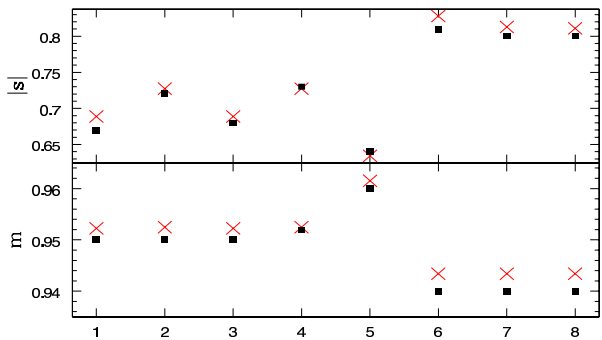}
  \end{center}
  \caption{Equal-mass BBHs with equal spin magnitudes and generic spin
    orientations.  Square data points 1 through 8 indicate the final
    kick $|{\bf k}|$, spin $|{\bf s}|$, and mass $m$ for eight
    simulations listed from left to right in Table I of
    \cite{Tichy:2007hk}.  Red X's show our predictions for these final
    quantities.}
  \label{Case5fig}
\end{figure}

Finally we examine the case of arbitrarily oriented initial spins.
From each of the 8 simulations in \cite{Tichy:2007hk} we have $|{\bf
  s}|$ and $m$.  We use the coefficients obtained from our fits in
Fig.~\ref{Case1fig} for the terms that depend only on $a_{3}$ and
$b_{3}$.  Then our 1st-order fits for $|{\bf s}|$ (with 3 free
parameters) and $m$ (with {\it zero} free parameters!) are shown in
Fig.~(\ref{Case5fig}).  We stress that this data set is not described
by any previous fitting formula.  The spin expansion gives the first
explanation for the distribution of points Fig.~\ref{Case5fig}.

In this {\it Letter}, we introduce a new {\it spin expansion} of final
quantities $f$ in triad components $a_{i}, b_{i}$ of the initial
spins.  Using the transformation properties of $f$, $a_{i}$, $b_{i}$
and $q$ under parity $P$ and exchange $X$, we dramatically reduce the
number of terms that one might naively expect.  Without resorting to
the sophisticated machinery of numerical relativity and the
post-Newtonian expansion, we obtain some detailed (and often new)
quantitative understanding of the final state of BBH merger.  This
clarifies the separation between the non-linear dynamics of Einstein's
equations and our more elementary non-dynamical considerations.

Our approach complements both post-Newtonian approximations and
numerical relativity.  Post-Newtonian methods provide accurate
predictions in the weak-field inspiral regime, but break down during
the later stages of the merger.  By contrast, the symmetries under $P$
and $X$ implicitly hold through the entire merger.  

Only numerical relativity can model the late stages of the merger, but
simulations remain computationally expensive.  The spin expansion
offers enormous computational savings in mapping the 7-dimensional
parameter space of BBH initial states $\{q, a_{i}, b_{i}\}$.  Even 10
grid points along each direction would mean $10^7$ simulations --- a
hopelessly large number.  However, the values $\{m, k_{i}^{},
s_{i}^{}\}$ from 16 independent simulations determine the spin
expansion coefficients up to 2nd order at fixed mass ratio $q$.  Then,
with 10 grid points for $q$, a mere 160 simulations could map the
space $\{q,a_{i},b_{i}\}$.  Further reductions are possible if the $q$
dependence of our coefficients can be identified analytically.  The
spin expansion may also help in identifying systematic errors with
forbidden geometrical dependence on the initial spins.

The spin expansion is useful for astrophysics and cosmology --- {\it
  e.g.}\ allowing BBH results from numerical relativity to be
efficiently included in simulations of cosmological structure
formation or black hole growth.  These simulations cannot resolve the
short scales relevant to supermassive BBH merger, and must instead
rely on maps from the initial to final state such as those presented
in this {\it Letter}.  The spin expansion predicts these final
quantities for {\it arbitrary} initial spin configurations, specified
once gravitational radiation (rather than dynamical friction)
dominates the binary evolution.

{\bf Acknowledgements.}  We thank Alessandra Buonanno, Neal Dalal,
Larry Kidder, Luis Lehner, Eric Poisson, Harald Pfeiffer, Jonathan
Sievers, Bill Unruh and Daniel Wesley for helpful conversations.

\end{document}